\documentclass{emulateapj}\usepackage{timesfonts}
\shorttitle{Diffusive Shock Acceleration}
\shortauthors{Kang et al.}
\slugcomment{draft of \today}

\def\etal{{\it et al.}}

\def\kms{~{\rm km~s^{-1}}}
\def\cm3{~{\rm cm^{-3}}}

\begin{document}
\title{Injection of $\kappa$-like Suprathermal Particles into Diffusive Shock
Acceleration}

\author{Hyesung Kang$^1$, Vah\'e Petrosian$^2$, Dongsu Ryu$^3$ and T. W.
Jones$^4$}
\affil{$^1$Department of Earth Sciences, Pusan National University, Pusan
609-735, Korea: hskang@pusan.ac.kr\\
$^2$Departments of Physics and Applied Physics, and KIPAC, Stanford University, 
Stanford, CA 94305, USA: vahe@stanford.edu\\
$^3$Department of Physics, UNIST, Ulsan 689-798, Korea: ryu@canopus.cnu.ac.kr\\
$^4$School of Physics and Astronomy, University of Minnesota, Minneapolis, MN
55455, USA: twj@msi.umn.edu}

\begin{abstract}

We consider a phenomenological model for the thermal leakage injection in the
diffusive shock
acceleration (DSA) process, in which suprathermal protons and electrons near the
shock transition zone
are assumed to have the so-called $\kappa$-distributions
produced by interactions of background  thermal particles with
pre-existing and/or self-excited plasma/MHD waves or turbulence.
The $\kappa$-distribution has a power-law tail, instead of an exponential
cutoff,
well above the thermal peak momentum.
So there are a larger number of potential seed particles with  momentum,
above that required for  participation in the DSA process. 
As a result, the injection fraction for the $\kappa$-distribution depends on the
shock Mach 
number much less severely compared to that for the Maxwellian distribution.
Thus, the existence of $\kappa$-like suprathermal tails at shocks would ease the
problem of extremely
low injection fractions, especially for electrons and especially at weak shocks
such as those found in the intracluster medium.
We suggest that the injection fraction for protons ranges $10^{-4}-10^{-3}$ for
a 
$\kappa$-distribution with $10 \la \kappa_p \la 30$ at quasi-parallel shocks,
while the injection fraction for electrons becomes $10^{-6}-10^{-5}$ for 
a $\kappa$-distribution with $\kappa_e \la 2$ at quasi-perpendicular shocks.
For such $\kappa$ values the ratio of cosmic ray electrons to protons naturally
becomes
$K_{e/p}\sim 10^{-3}-10^{-2}$,
which is required to explain the observed ratio for Galactic cosmic rays.

\end{abstract}


\section{Introduction}

Acceleration of nonthermal particles is ubiquitous at astrophysical
collisionless shocks,
such as interplanetary shocks in the solar wind,
supernova remnant (SNR) shocks in the interstellar medium (ISM) and structure
formation
shocks in the intracluster medium (ICM) \citep{blaeic87,jones91,ryuetal03}.
Plasma physical processes operating at collisionless shocks, 
such as excitation of waves via plasma instabilities and
ensuing wave-particle interactions, depend primarily on the shock magnetic field
obliquity as well as 
on the sonic and Alfv\'enic Mach numbers, $M_s$ and $M_A$, respectively.
Collisionless shocks can be classified into two categories by the obliquity
angle, $\Theta_{BN}$,
the angle between the upstream mean magnetic field and the shock normal:
quasi-parallel ($\Theta_{BN} \la 45^{\circ}$) and quasi-perpendicular
($\Theta_{BN} \ga 45^{\circ}$).
Diffusive shock acceleration (DSA) at strong SNR shocks with $M_s \sim M_A \sim
10-100$ is reasonably 
well understood, especially for the quasi-parallel regime,
and it has been tested via radio-to-$\gamma$-ray observations of nonthermal
emissions from 
accelerated cosmic ray (CR) protons and electrons \citep[see][for
reviews]{dru83, blaeic87,hill05,reynolds12}.
On the contrary, DSA at weak shocks in the ICM ($M_s\sim 2-3$, $M_A\sim 10$)
is rather poorly understood, although its signatures have apparently been
observed in a number of radio
relic shocks \citep[e.g.][]{wrbh10,feretti12,krj12,brunetti2014}.
At the same time, {\it in situ} measurements of Earth's bow shock, or traveling
shocks
in the interplanetary medium (IPM) with spacecrafts have
provided crucial insights and tests for plasma physical processes related with
DSA
at shocks with moderate Mach numbers ($M_s\sim M_A \la 10$)
\citep[e.g.][]{shimada99,oka06,zank07,masters13}.

Table 1 compares characteristic parameters for plasmas in the IPM, ISM (warm
phase), and ICM
to highlight their similarities and differences.
Here the plasma beta $\beta_p~(=P_g/P_B\propto n_H T/B_0^2$) is the ratio of the
thermal to magnetic pressures,
so the magnetic field pressure is dynamically more important in lower beta
plasmas.
The plasma alpha is defined as the ratio of the electron plasma
frequency to cyclotron frequency:
\begin{equation}
\alpha_p~= {\omega_{pe}\over \Omega_{ce}} = \frac{2\pi r_{ge}}{\lambda_{De}}
\approx {{\sqrt{m_e/m_p}\cdot c} \over v_A}
\propto {\sqrt{n_e}\over B_0},
\end{equation}
where $r_{ge}$ is the electron gyroradius, and $\lambda_{De}$ is the electron
Debye length, $v_A=B_0/\sqrt{4\pi \rho}$ is the Alfv\'en speed.
Plasma wave-particle interactions and ensuing stochastic acceleration 
are more significant in lower alpha plasmas \citep[e.g.][]{pryadko97}.
Among the three kinds of plasmas in Table 1, the ICM with the highest $\beta_p$
has dynamically least
significant magnetic fields,
but, with the smallest $\alpha_p$, plasma interactions are expected to be most
important there.
The last three columns of Table 1 show typical shock speeds, sonic Mach numbers,
and Alfv\'enic Mach numbers for interplanetary shocks near 1 AU,
SNR shocks and ICM shocks.

This paper focuses on the injection of suprathermal particles into the DSA
process
at astrophysical shocks.
Since the shock thickness is of the order of the gyroradius of postshock
thermal protons,
only suprathermal particles (both protons and electrons) with momentum
$p\ga p_{\rm inj}\approx (3-4) p_{\rm th,p}$ can re-cross to the shock upstream
and participate 
in the DSA process \citep[e.g.][]{kjg02}.
Here, $p_{\rm th,p}=\sqrt{2m_p k_B T_2}$ is the most probable momentum of
thermal protons with 
postshock temperature $T_2$ and  $k_B$ is the Boltzmann constant.
Hereafter, we use the subscripts `1' and `2' to denote the
conditions upstream and downstream of shock, respectively.
At quasi-parallel shocks, in the so-called {\it thermal injection leakage}
model,
protons leaking out of the {\it postshock} thermal pool are assumed to
interact 
with magnetic field fluctuations and become the CR population
\citep[e.g.][]{maldru01,kjg02}. 
In a somewhat different interpretation based on hybrid plasma simulations,
protons
reflected off the shock transition layer are thought to form a beam of streaming
particles, 
which in turn excite resonant waves that scatter particles into the DSA process
\citep[e.g.][]{quest88,guo13}.
At quasi-perpendicular shocks, on the other hand, the self-excitation of waves
is ineffective and
the injection of suprathermal protons is suppressed significantly
\citep{capri13},
unless there exists pre-existing MHD turbulence in the background plasma
\citep{giacalone05,zank06}.

Assuming that downstream electrons and protons have the same kinetic temperature
($T_e\approx T_p$), for a Maxwellian distribution there will
be fewer electrons than protons that will have momenta above the required
injection momentum. Thus 
electrons must  be pre-accelerated from the thermal momentum ($p_{\rm
th,e}=(m_e/m_p)^{1/2} p_{\rm th,p}$)
to the injection momentum ($p_{\rm inj}\approx (130-170) p_{\rm th,e}$) 
in order to take part in the DSA process.
Contrary to the case of protons, which are effectively injected at
quasi-parallel
shocks, according to {\it in situ} observations made by spacecrafts, electrons
are known to be
accelerated at Earth's bow shock and interplanetary shocks {\it preferentially}
in  the quasi-perpendicular configuration
\citep[e.g.][]{gosling89,shimada99,simnett05,oka06}.
However, in a recent observation of Saturn's bow shock by the Cassini
spacecraft, 
the electron injection/acceleration has been detected also in the quasi-parallel
geometry
at high-Mach, high-beta shocks ($M_A\sim 100$ and $\beta_p\sim 10$)
\citep{masters13}.
\citet{riqu11} suggested that electrons can be  injected and accelerated also at
quasi-parallel portion of
{\it strong} shocks such as SNR shocks, 
because the turbulent magnetic fields excited by the CR streaming instabilities
upstream of the shock may have perpendicular components
at the corrugated shock surface. 
So, locally transverse magnetic fields near the shock surface
seem essential for the efficient electron injection regardless of the obliquity
of the large-scale,
mean field.

Non-Maxwellian tails of high energy particles have been widely observed in
space and laboratory plasmas \citep[e.g.][]{vasyliunas68,hellberg00}.
Such particle distributions can be described
by the combination of a Maxwellian-like core and a suprathermal tail of
power-law form,
which is known as the $\kappa$-distribution.
There exists an extensive literature that explains the $\kappa$-distribution from
basic physical
principles and processes relevant for collisionless, weakly coupled plasmas
\citep[e.g.][]{leubner04,pierrard10}.
The theoretical justification for the $\kappa$-distribution is beyond the scope
of this paper,
so readers are referred to those papers.
Recently, the existence of $\kappa$-distribution of electrons
has been conjectured and examined in order to explain the discrepancies 
in the measurements of electron temperatures and metallicities 
in H II regions and planetary nebulae \citep{nicholls12,mendoza14}. 

The development of suprathermal tails of both proton and electron distributions
are 
two outstanding problems in the theory of collisionless shocks, 
which involve complex wave-particle interactions such as
the excitation of kinetic/MHD waves via plasma instabilities and the
stochastic acceleration by plasma turbulence \citep[see][for recent
reviews]{petrosian12,schure12}.
For example, stochastic acceleration of thermal electrons by electron-whistler
interactions 
is known to be very efficient in low $\beta_p$ and low $\alpha_p$ plasmas such
as
solar flares \citep{hamilton92}.
Recently, the pre-heating of electrons and the injection of protons at
non-relativistic 
collisonless shocks have been studied using Particle-in-Cell (PIC) and hybrid
plasma 
simulations for a wide range of parameters
\citep[e.g.][]{amano09,guo10,guo13,riqu11,garat12,capri13}. 
In PIC simulations, the Maxwell's equations for electric and magnetic fields are
solved 
along with the equations of motion for ions and electrons, so
full wave-particle interactions can be followed from first principles.
In hybrid simulations, only ions are treated kinetically,
while electrons are treated as a neutralizing, massless fluid.

Using two and three-dimensional PIC simulations, 
\citet{riqu11} showed that for low Alfv\'enic Mach numbers ($M_A \la 20$),
oblique whistler waves can be excited in the foot of quasi-perpendicular shocks
(but not at perfectly perpendicular shocks with $\Theta_{Bn}=90^{\circ}$).
Electrons are then accelerated via wave-particle interactions with those
whistlers,
resulting in a power-law suprathermal tail.
They found that the suprathermal tail can be represented by the energy spectrum
$n_e(E)\propto E^{-a}$
with the slope $a=3-4$, which is harder for smaller $M_A$ (i.e., larger $v_A$ or
smaller $\alpha_p$).
Nonrelativistic electrons streaming away from a shock
can resonate only with high frequency whistler waves with right hand helicity,
while  protons and relativistic electrons (with Lorentz factor $\gamma>m_p/m_e$)
resonate with MHD (Alfv\'en) waves.
So the generation of oblique whistlers is thought to be one of 
the agents for pre-acceleration of electrons \citep{shimada99}. 
In fact, obliquely propagating whistler waves and high energy electrons are
often observed together in the upstream region of quasi-perpendicular
interplanetary shocks \citep[e.g.][]{shimada99, wilson09}.
Recently, \citet{wilson12} observed obliquely propagating whistler modes in the
precursor 
of several quasi-perpendicular interplanetary shocks with low Mach numbers 
(fast mode Mach number $M_f\approx 2-5$), 
simultaneously with perpendicular ion heating and parallel electron
acceleration.
This observation implies that oblique whistlers could play an important role in
the development of a
suprathermal halo around the thermal core in the electron velocity distribution
at quasi-perpendicular shocks with moderate $M_A$.

Using two-dimensional PIC simulations
for perpendicular shocks with $M_A\sim 45$, \citet{matsumoto13} found that
several kinetic
instabilities (e.g. Buneman, ion-acoustic, ion Weibel) are excited at the
leading edge of the shock foot 
and that electrons can be energized to relativistic energies via the shock
surfing mechanism.
They suggested that the shock surfacing acceleration can provide the effective
pre-heating of
electrons at strong SNR shocks with high Alfv\'enic Mach numbers ($M_A\ga 100$).

Because non-relativistic electrons and protons interact with different types of
plasma waves
and instabilities, they can have suprathermal tails with different properties
that depend on
plasma and shock parameters, such as $\Theta_{Bn}$, $\alpha_p$, $\beta_p$,
$M_s$, and $M_A$.
So the power-law index of the $\kappa$-distributions for electrons and protons, 
$\kappa_e$ and $\kappa_p$, respectively, should depend on these parameters,
and they could be significantly different from each other.
For example, the electron distributions measured in the IPM can be fitted with 
the $\kappa$-distributions with $\kappa_e\sim 2-5$, 
while the proton distributions prefer a somewhat larger $\kappa_p$
\citep{pierrard10}.
Using {\it in situ} spacecraft data, \citet{parker12} suggested that the proton
spectra observed
downstream of {\it quasi-parallel} interplanetary shocks 
can be explained by the injection from the upstream (solar-wind) thermal
Maxwellian or 
weak $\kappa$-distribution with $\kappa_p\ga10$.
On the other hand, \citet{parker14} showed that the upstream suprathermal tail 
of the $\kappa_p=4$ distribution is the best to fit the proton spectra observed
downstream 
of {\it quasi-perpendicular} interplanetary shocks\footnote{Note that
\citet{parker12} and \citet{parker14} model particles 
from the {\it upstream} suprathermal pool being injected into the DSA process, 
while here we assume that particles from the {\it downstream} suprathermal pool
are injected.}.
They reasoned that the upstream proton distribution may form a relatively flat
$\kappa$-like suprathermal tail due to
the particles reflected at the magnetic foot of quasi-perpendicular shocks,
while at quasi-parallel shocks the upstream proton distribution remains
more-or-less Maxwellian.

In this paper, we consider a phenomenological model for the thermal leakage
injection in the DSA process
by taking the $\kappa$-distributions as empirical forms for the suprathermal
tails of the
electron and proton distributions at collisionless shocks.
The $\kappa$-distribution is described in Section 2.
The injection fraction is estimated in Section 3,
followed by a brief summary in Section 4.

\section{Basic Models}

For the postshock nonrelativistic gas of kinetic
temperature $T_2$ and particle density $n_2$,
the Maxwellian momentum distribution is given as
\begin{equation}
f_{\rm M}(p)= {n_2 \over \pi^{1.5}}~ p_{\rm th}^{-3} ~\exp\left[-\left({p\over
p_{\rm th}}
\right)^2\right],
\label{fmaxw}
\end{equation}
where $p_{\rm th}= \sqrt{2 m k_B T_2}$ is the thermal peak momentum
and the mass of the particle is $m=m_e$ for electrons and $m=m_p$ for protons.
The distribution function is defined in general as $\int 4\pi p^2f(p) dp = n_2$.
Here we assume that the electron and proton distributions have the same kinetic
temperature, 
so that $p_{\rm th,e}= \sqrt{m_e/m_p} \cdot p_{\rm th,p}$.

The $\kappa$-distribution can be described as
\begin{equation}
f_{\kappa}(p)= {n_2 \over \pi^{1.5}}~ p_{\rm th}^{-3} { {\Gamma(\kappa+1) }
\over {(\kappa-3/2)^{3/2}\Gamma(\kappa-1/2) } }
\left[1+{p^2\over {(\kappa-3/2)p_{\rm th}^2}}\right]^{-(\kappa+1)},
\label{fkappa}
\end{equation}
where $\Gamma(x)$ is the Gamma function
\citep[e.g.][]{pierrard10}.
The $\kappa$-distribution asymptotes to a power-law form, $f_{\kappa}(p)\propto
p^{-2(\kappa+1)}$
for $p\gg p_{\rm th}$,
which translates into $N(E)\propto E^{-2\kappa}$ for relativistic energies,
$E\ga mc^2$.
For large $\kappa$, it asymptotes to the Maxwellian distribution.
For a smaller value of $\kappa$, the $\kappa$-distribution has a flatter,
suprathermal, power-law tail,
which may result from larger wave-particle interaction rates.
Note that for the $\kappa$-distribution in equation (\ref{fkappa}), the mean
energy per particle, 
$m \langle v^2 \rangle/2= (2\pi m/n_2) \int v^2 f_{\kappa}(p)  p^2 dp$,
becomes $(3/2)k_B T_2$ and the gas pressure 
becomes $P_2= n_2 k_B T_2$, providing that particle speeds are nonrelativisitic.

The top panel of Figure 1 compares $f_{\rm M}$ and $f_{\kappa}$ for 
electrons and protons when $T_2 = 5\times 10^7$ K 
(corresponding to the shock speed of $u_s\approx 1.9\times10^3\kms$ in the large
$M_s$ limit.) 
Here, the momentum is expressed in units of $m_e c$ for both electrons and
protons,
so the distribution function $f(p)$ is plotted in units of $n_2/(m_e c)^3$.
Note that the plotted quantity is $p^3 f(p) d\ln p = p^2 f(p)dp \propto n(p)dp
$.
For smaller values of $\kappa$, the low energy portion of $f_{\kappa}(p)$ also
deviates more significantly from $f_{\rm M}(p)$.

For the $\kappa$-distribution, the {\it most probable} momentum (or the peak momentum)
is related to the Maxwellian peak momentum as
$p_{\rm mp}^2 = p_{\rm th}^2 \cdot (\kappa-3/2)/\kappa$.
So for a smaller $\kappa$, the ratio of $p_{\rm mp}/p_{\rm th}$ becomes smaller.
In other words,
the peak of $f_{\kappa}(p)$ is shifted to a lower momentum for a smaller
$\kappa$, 
as can be seen in the top panel of Figure 1.
To account for this we will suppose a hypothetical case in which the postshock temperature is
modified 
for a $\kappa$-distribution as follows:
\begin{equation}
T_2^{\prime}(\kappa) = T_2 {\kappa \over {(\kappa-3/2)} }.
\label{Tprime}
\end{equation}
Then the most probable momentum becomes the same for different $\kappa$'s.
The bottom panel of Figure 1 compares the Maxwellian distribution for $T_2 =
5\times 10^7$ K and
the $\kappa$-distributions with the corresponding $T_2^{\prime}(\kappa)$'s.
For such $\kappa$-distributions,
the distribution of low energy particles with $p\la p_{\rm th}$
remains very similar to the Maxwellian distribution.
In that case, low energy particles follow more-or-less the Maxwellian
distribution, while
higher energy particles above the thermal peak momentum show a power-law tail.
This might represent the case in which thermal particles with $p\ga p_{\rm th}$ 
gain energies via stochastic acceleration by pre-existing
and/or self-excited waves in the shock transition layer, resulting in a
$\kappa$-like tail and
additional plasma heating.
Such $\kappa$-distributions with plasma heating could be close to the real
particle
distributions behind collisionless shocks.
So below we will consider two cases: the $T_2$ model in which the postshock
temperature is
same and the $T_2^{\prime}$ model in which the postshock temperature depends on
$\kappa$ as
in equation (\ref{Tprime}).

\section{Injection Fraction}
 
We assume that the distribution function of the particles accelerated by DSA,
which we refer to as cosmic rays (CRs), at the position of the shock has the test-particle
power-law spectrum
for $p\ge p_{\rm inj} \equiv Q_{\rm inj} \cdot p_{\rm th,p}$,
\begin{equation}
f_{\rm CR}(p)=f(p_{\rm inj})\cdot \left({p \over p_{\rm inj}}\right)^{-q},
\label{ftp}
\end{equation}
where the power-law slope is given as
\begin{equation}
q={{3(u_1-v_{A,1})}\over u_1-v_{A,1}-u_2}.
\label{qtp}
\end{equation}
Here $u_1$ and $u_2$ are the upstream and downstream flow speeds, respectively,
in the shock rest frame, and $v_{A,1}=B_1/\sqrt{4\pi\rho_1}$ is the upstream
Alfv\'en speed.
This expression takes account of the drift of the Alfv\'en waves excited by
streaming
instabilities in the shock precursor \citep[e.g.,][]{kang11, kang12}.
If $v_{A,1}=0$, the power-law slope becomes $q=4$ for $M_s\gg1$ and $q=4.5$ for
$M_s=3$.

Note that in our phenomenological model, we assume the $\kappa$-distribution
extends only to $p=p_{\rm inj}$,
above which the DSA power-law in equation (\ref{ftp}) sets in.
In Figure 1 the vertical dotted lines show the range of $p_{\rm inj} = (3.5-4)\
p_{\rm th,p}$,
above which the particles can participate the DSA process.
With $\kappa_p=30$ for protons, $\kappa_e=2$ for electrons, and $p_{\rm inj}=4
p_{\rm th,p}$,
for example,
the ratio of $f_e(p_{\rm inj})/f_p(p_{\rm inj})\approx 10^{-2.6}$ for the $T_2$
model,
while $f_e(p_{\rm inj})/f_p(p_{\rm inj})\approx 10^{-1.9}$ for the
$T_2^{\prime}$ model.

The parameter $Q_{\rm inj}$ determines
the CR injection fraction, $\xi \equiv {n_{CR} / n_2 }$ as follows.
In the case of the Maxwellian distribution the fraction is
\begin{equation}
\xi_{\rm M}  = {4 \over \sqrt{\pi}} 
{ {Q_{\rm inj}^3} \over {(q - 3)}}\cdot \exp(-Q_{\rm inj}^2) ,
\label{xiM}
\end{equation}
while in the case of the $\kappa$-distribution it is
\begin{equation}
\xi_{\kappa} = {4 \over \sqrt{\pi}} 
{{Q_{\rm inj}^3}  \over {(q - 3)}} \cdot
{ {\Gamma(\kappa+1) } \over {(\kappa-3/2)^{3/2}\Gamma(\kappa-1/2) } }
\left[1+{Q_{\rm inj}^2\over {(\kappa-3/2)}}\right]^{-(\kappa+1)}.
\label{xiK}
\end{equation}
Note that both forms of the injection fraction are independent of the postshock
temperature
$T_2$, but dependent on $Q_{\rm inj}$ and the shock Mach number, through the
slope $q(M_s)$.

For the Maxwellian distribution, 
$\xi_{\rm M}$ decreases exponentially with the parameter $Q_{\rm inj}$,
which in general depends on the shock Mach number as well as on the obliquity.
Since the injection process should depend on the level
of pre-existing and self-excited plasma/MHD waves,
$Q_{\rm inj}$ is expected to increase with $\Theta_{Bn}$.
For example, in a model adopted for quasi-parallel shocks \citep[e.g.][]{kr10},
\begin{equation} 
Q_{\rm inj}\approx \chi { {m_p u_2}\over {p_{\rm th,p}}}
= \chi \sqrt{{\gamma\over{2\mu}}}{u_2 \over c_{s,2}}=  \chi
\sqrt{{\gamma\over{2\mu}}}
\left[{ { (\gamma-1)M_s^2+2} \over {2\gamma M_s^2-(\gamma-1)} }\right]^{1/2},
\label{Qinj}
\end{equation}
where $\chi \approx 5.8-6.6$, $\gamma$ is the gas adiabatic index,
and $\mu$ is the mean molecular weight for
the postshock gas.
For $\gamma=5/3$ and $\mu=0.6$, this parameter approaches to $Q_{\rm inj}\approx
3-4$ for large $M_s$,
depending on the level of MHD turbulence, 
and it increases as $M_s$ decreases (see Figure 1 of \citet{kr10}).
Using hybrid plasma simulations, 
\citet{capri13} suggested $Q_{\rm inj}=3-4$ at quasi-parallel shocks with
$M_s\approx M_A\approx 20$,
leading to the injection fraction of $\xi_p\approx 10^{-4}-10^{-3}$ for protons.

For highly oblique and perpendicular shocks, the situation is
more complex and  the modeling of $Q_{\rm inj}$  becomes difficult,
partly because MHD waves are not self-excited effectively and
partly because the perpendicular diffusion is not well understood
\citep[e.g.][]{parker14}.
So the injection process at quasi-perpendicular shocks depends on the
pre-existing MHD turbulence
in the upstream medium as well as the angle $\Theta_{Bn}$.
For example, \citet{zank06} showed that 
in the case of interplanetary shocks in the solar wind
located near 1AU from the sun,
the injection energy is similar for $\Theta_{Bn}=0^{\circ}$ and $90^{\circ}$,
but it peaks at highly oblique shocks with $\Theta_{Bn}\sim 60-80^{\circ}$.
So $Q_{\rm inj}$ would increase with $\Theta_{Bn}$, but decrease as
$\Theta_{Bn}\rightarrow 90^{\circ}$.
The same kind of trend may apply for cluster shocks, but again the details will
depend on the MHD turbulence in
the ICM.
Here we will consider a range of values, $3\le Q_{\rm inj} \le 5$.
For the $\kappa$-distribution, $\xi_{\kappa}$ also decreases with $Q_{\rm inj}$
but more
slowly than $\xi_M$ does.
This means that the dependence of injection fraction on the shock sonic
Mach number would be weaker in the case of the $\kappa$-distribution.

Figure 2 shows the energy spectrum of protons for the two (i.e., $T_2$ and
$T_2^{\prime}$) models shown in Figure 1.
Here the energy spectrum is calculated as
$n_p(E)= 4\pi p^2 f(p) ({dp}/{dE})$,
where the kinetic energy is $E=\sqrt{p^2c^2+m_p^2c^4} - m_p c^2$
and the distribution function $f(p)$ is given in equations (\ref{fmaxw}) or
(\ref{fkappa}).
The filled and open circles mark the spectrum at the energies corresponding to
$3.5\ p_{\rm th,p}$ and $4\ p_{\rm th,p}$ for the Maxwellian distribution
and the $\kappa$-distributions with $\kappa_p= 10$ and 30.
This shows that the injection efficiency for CR protons would be enhanced in
the $\kappa$-distributions, compared to the Maxwellian distribution,
by a factor of $\xi_{\kappa_p=10}/\xi_{\rm M}\sim 100-300$ 
and $\xi_{\kappa_p=30}/\xi_{\rm M}\sim 10-20$.

There are reasons why the cases of $\kappa_p=10-30$ are shown here.
It has been suggested that the upstream suprathermal populations can be
represented by
the $\kappa$-distribution with $\kappa_p\approx 4$ at quasi-perpendicular IPM
shocks \citep{parker14}
and $\kappa_p\ga 10$ at quasi-parallel IPM shocks \citep{parker12}.
However, the proton injection at quasi-parallel shocks is much more efficient
than that
at quasi-perpendicular shocks,
because the injection energy is much higher at highly oblique shocks
\citep[e.g.][]{zank06}.
Moreover, \citet{capri13} showed that the proton injection at quasi-parallel
shocks 
can be modeled properly with the thermal leakage injection from the Maxwellian
distribution 
at $p_{\rm inj}\approx (3-4)p_{\rm th,p}$.
They also showed that a harder suprathermal population forms at larger
$\Theta_{BN}$,
which is consistent with the observations at IPM shocks.
But the power-law CR spectrum does not develop at (almost)
perpendicular shocks due to lack of self-excited waves in their hybrid
simulations.

As shown in Figure 1 the electron distribution needs a substantially
more enhanced suprathermal tail,
for example, the one in the $\kappa$-distribution with $\kappa_e\sim 2$,  
in order to achieve the electron-to-proton ratio $K_{e/p}\sim 10^{-3}-10^{-2}$
with the thermal leakage injection model.
Figure 3 shows the energy spectrum of electrons for the two models shown in
Figure 1.
Here the energy spectrum for electrons is calculated as
$n_e(\Gamma_e -1 )= 4\pi p^2 f(p) ({dp}/{d\Gamma_e})$,
where the Lorentz factor is $\Gamma_e=\sqrt{1+ (p/m_e c)^2}$.
The filled and open circles mark the spectrum at the energies corresponding to
$p_{\rm inj}= (3.5-4)\ p_{\rm th,p}$ for the $\kappa$-distributions
with $\kappa_e= 1.6, 2.0$, and 2.5.
Note that the $\kappa$-distribution is defined for $\kappa>3/2$.
In the PIC simulations of quasi-perpendicular shocks by \citet{riqu11},
the power-law slope of $n_e(E)$ at $\Gamma_e\sim 10-100$ ranges $2.7<a<4$ for $
3.5\le M_A\le 14$,
where $m_p/m_e=1600$ was adopted (see their Figure 12).
This would translate roughly into $\kappa_e \la 2$, which is consistent with the
observations at quasi-perpendicular IPM shocks \citep{pierrard10}.

If the suprathermal tails of electrons and protons can be described by the
$\kappa$-distributions with $\kappa_e$ and $\kappa_p$, respectively,  for $p\le
p_{\rm inj}$,
and if both CR electrons and protons have simple power-laws given in equation
(\ref{ftp})
for $p> p_{\rm inj}$,
then the injection fractions, $\xi_p$ and $\xi_e$, for $\kappa$-distributions 
can be estimated by equation (\ref{xiK}).
Figure 4 compares the injection fractions, $\xi_p$ and 
$\xi_e $, for the two models shown in Figures 1-3.
Note that the slope $q$ depends on $M_s$, so $\xi (q-3)$ is plotted instead of just $\xi$.
Now the ratio of CR electron to proton numbers can be calculated as
\begin{equation}
K_{e/p}(Q_{\rm inj},\kappa_e,\kappa_p )\equiv {{\xi_e(Q_{\rm inj},\kappa_e)}
\over 
{\xi_p(Q_{\rm inj},\kappa_p)}}={{f_e(p_{\rm inj},\kappa_e)}\over {f_p(p_{\rm
inj},\kappa_p)}}.
\end{equation}

In the $\kappa$-distribution of protons with $\kappa_p=30$ (dot-dashed line),
for example, 
$\xi_p$ decrease from $10^{-3}$ to $10^{-4}$ when $Q_{\rm inj}$ increase from
3.5 to 4.
We note that for $\kappa_p\la 10$ or for $Q_{\rm inj}\la 3.5$,
the proton injection fraction would be too high (i.e., $\xi_p>10^{-3}$)
to be consistent with commonly-accepted DSA modelings of observed shocks such as
SNRs.
The parameter $Q_{\rm inj}$ would in general increase for a smaller $M_s$ as
illustrated in
equation (\ref{Qinj}).
The dependence of the injection fraction on $M_s$ becomes weaker 
for the $\kappa$-distribution than for the Maxwellian distribution.
As a result, the suppression of the CR injection fraction at weak shocks will be
less severe
if the $\kappa$-distribution is considered.

For electrons, the injection fraction would be too small if they were to be
injected by way of thermal leakage from the Maxwellian distribution.
So that case is not included in Figure 4.
The expected electron injection would be $\xi_e\sim 10^{-6}-10^{-5}$,
if one takes $K_{e/p}\sim 10^{-3}-10^{-2}$ and $\xi_p\sim 10^{-4}-10^{-3}$.
Then, the suprathermal tails of the $\kappa$-distributions with $\kappa_e\la 2$
would be necessary.
For electron distributions with such flat suprathermal tails,
the injection fraction would not be significantly suppressed even at weak
shocks.

Turbulent waves excited in the shock precursor/foot should decay away from the shock (both
upstream and downstream), as seen in the interplanetary shocks \citep{wilson12} 
and the PIC simulations \citep{riqu11}. 
Thus it is possible the $\kappa$-like suprathermal electron populations exist only in a narrow region 
around the shock, and in any case the differences from Maxwellian form that we discuss here are too 
limited to produce easily observable signatures such as 
clearly nonthermal hard X-ray bremsstrahlung.

During the very early stage of SNR expansion with $u_s > 10^4 \kms$, 
the postshock electrons should be described by the relativistic Maxwellian distribution
with a relatively slow exponential cutoff of $\exp(-\Gamma_e m_e c^2/k_B T)$ 
instead of equation (\ref{fmaxw}).
However, injection from relativistic electron plasmas at collisionless shocks could 
involve much more complex plasma processes and lie beyond the scope of this study.

\section{Summary}

In the so-called thermal leakage injection model for DSA,
the injection fraction depends on the number of suprathermal particles near
the injection momentum, $p_{\rm inj}=Q_{\rm inj} p_{\rm th,p}$, 
above which the particles can participate in the 
DSA process \citep[e.g.][]{kjg02}.
The parameter $Q_{\rm inj}$ should be larger for larger oblique angle,
$\Theta_{Bn}$, 
and for smaller sonic Mach number, $M_s$, leading to a smaller injection
fraction.
Moreover, it should depend on the level of magnetic field turbulence, both
pre-existing and self-excited,
which in turn depends on the plasma parameters such as $\beta_p$ and $\alpha_p$
as well as
the power-spectrum of MHD turbulence.
Since the detailed plasma processes related with the injection process are not
fully understood,
here we consider a feasible range, $3\le Q_{\rm inj}\le 5$. 

Assuming that suprathermal particles, both protons and electrons,
follow the $\kappa$-distribution with a wide range of the
power-law index, $\kappa_p$ and $\kappa_e$,
we have calculated the injection fractions for protons and electrons.
A $\kappa$-type distribution or distribution consisting of a 
quasi-thermal plus a nonthermal tail, with a short dynamic range as the one
needed here, is expected in a variety of models for acceleration of
nonrelativistic thermal particles (see e.g. \citet{petrosian08} for acceleration
in ICM or \citet{petrosian04} for acceleration in Solar flares).
The fact that efficient accelerations of electrons and protons require
$\kappa$-type distributions with different values of $\kappa$ suggests  that
they are produced by interactions with different types of waves; e.g., Alfven
waves for protons and whistler waves for electrons. 
We show  that $\kappa_p\sim 10-30$ leads to the injection fraction of
$\xi_p\sim 10^{-4}-10^{-3}$ for protons
at quasi-parallel shocks,
while $\kappa_e\la 2$ leads to the injection fraction of
$\xi_e \sim 10^{-6}-10^{-5}$ for electrons
at quasi-perpendicular shocks.
The proton injection is much less efficient at quasi-perpendicular shocks,
compared to quasi-parallel shocks, because MHD waves are not efficiently
self-excited
\citep{zank06,capri13}.
For electrons, a relatively flat $\kappa$-distribution may form due to obliquely
propagating
whistlers at quasi-perpendicular shocks with moderate Mach numbers ($M_A\la20$),
 and
$\kappa_e$ is expected to decrease for a smaller $M_A$ (i.e. smaller $\alpha_p$
or stronger magnetization) \citep{riqu11}.
We note that these $\kappa$-like suprathermal populations are expected to exist
only in a narrow region around the shock, since they should be produced via
plasma/MHD interactions with various waves, 
which could be excited in the shock precursor and then decay downstream.

In addition, we point out that acceleration (to high CR
energies) is less sensitive to  
shock and plasma parameters for a $\kappa$-distribution than the Maxwellian distribution.
So, the existence of $\kappa$-like suprathermal tails in the electron
distribution
would alleviate the problem of extremely low injection fractions for weak
quasi-perpendicular shocks
such as those widely thought to power radio relics found in the outskirts of
galaxy clusters \citep{krj12,pop13,brunetti2014}.

Finally, we mention that electrons are not likely to be accelerated at weak
quasi-parallel shocks, 
according to {\it in situ} measurements of interplanetary shocks
\citep[e.g.][]{oka06} and
PIC simulations \citep[e.g.][]{riqu11}.
At strong quasi-parallel shocks, on the other hand, \citet{riqu11} suggested
that electrons could be
injected efficiently through locally perpendicular portions of the shock
surface,
since turbulent magnetic fields are excited and amplified by CR protons
streaming ahead of the shock.
Thus the magnetic field obliquity, both global and local to the shock surface,
and magnetic field amplification via wave-particle interactions are among the key
players that govern 
the CR injection at collisionless shocks and need to be further studied by plasma
simulations.

\acknowledgements

HK thanks KIPAC for hospitality during the sabbatical leave
at Stanford University, where a part of work was done.
HK was supported by the National Research Foundation of Korea Grant funded 
by the Korean Government (NRF-2012-013-2012S1A2A1A01028560).
VP was supported
by NASA grants NNX10AC06G, NNX13AF79G and NNX12AO78G.
DR was supported by the National Research Foundation of Korea through grant
2007-0093860. TJ was supported by NSF grant AST1211595, NASA grant NNX09AH78G,
and the Minnesota Supercomputing Institute.
The authors would like to acknowledge the valuable comments
from an anonymous referee.

\begin{table*}
\begin{center}
{\bf Table 1.}~~Characteristic Plasma Parameters$^{\rm a}$\\
\vskip 0.3cm
\begin{tabular}{ lrrrrrrrrrr }
\hline\hline

 & $n_H$ & $T$ & {$B$}& {$c_s$}& {$v_A$}&{~$\beta_p$}& {$\alpha_p$} & {$u_s$} & {$M_s$} & {$M_A$}\\

 {} &  {\small $(\rm cm^{-3})$} & {\small (K)} & {\small ($\mu$G)} & {\small (${\rm km~s^{-1}}$)} & {\small (${\rm
km~s^{-1}}$)} &
 {\small ($P_g/P_B$)}  &  ({\small $\omega_{pe}/\Omega_e$}) & {\small (${\rm km~s^{-1}}$)}  & & \\

\hline
IPM  & 5         & $10^5$        & 50  & 50     & 40 & 1.6 & 140 & $5\times 10^2$ & 10 &13\\
ISM  & 0.1       & $10^4$        & 5   & 15    & 30 & 0.3 & 200 & $3\times 10^3$ & 200 & 100 \\
ICM  & $10^{-4}$ & $5\times10^7$ & 1   & $10^3$ & 180 & 40 & 30 & $2\times 10^3$ & 2 & 11\\

\hline
\end{tabular}
\end{center}
$^{\rm a}${IPM=interplanetary medium, ISM=interstellar medium, ICM=intracluster
medium}\\

\end{table*}

\clearpage

\begin{figure}
\vspace{-1cm}
\begin{center}
\includegraphics[scale=0.85]{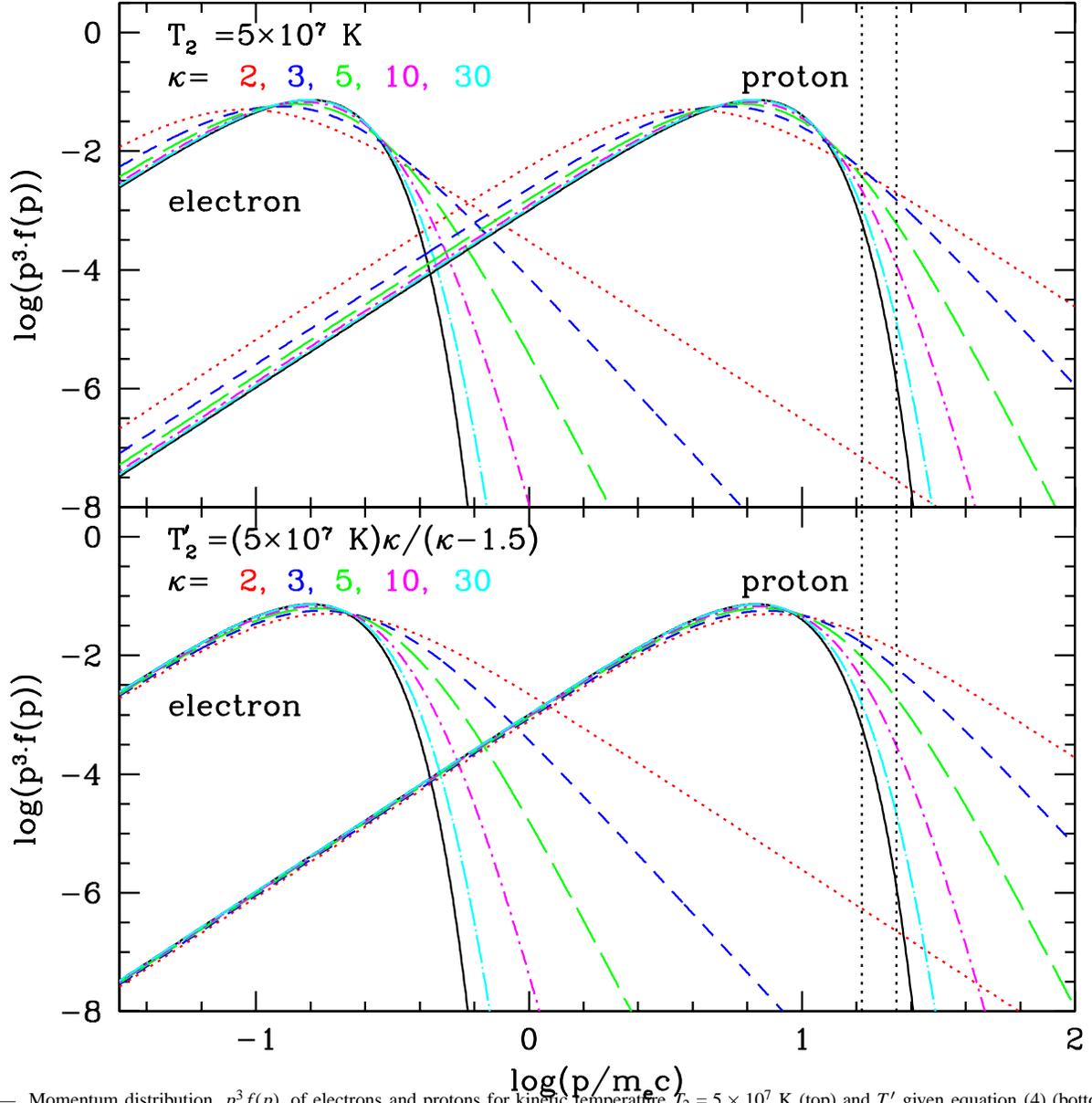}
\end{center}
\vspace{-1.0cm}
\caption{ 
Momentum distribution, $p^3f(p)$, of electrons and protons for kinetic
temperature 
$T_2=5\times 10^7$ K (top) and $T_2^{\prime}$ given equation (\ref{Tprime})
(bottom). 
The momentum is expressed in units of $m_ec$ for both electrons and protons,
and the distribution function $f(p)$ is plotted in units of $n_2/(m_e c)^3$.
The Maxwellian distributions are shown in (black) solid lines, while the
$\kappa$-distributions are shown in
(red) dotted, (blue) dashed, (green) long dashed, (magenta) dot-dashed, and
(cyan) dot-long dashed lines
for $\kappa=$ 2, 3, 5, 10, and 30, respectively.
The vertical lines indicate the range of the injection momentum of $p_{\rm inj}=
(3.5-4)\ p_{\rm th,p}$,
above which particles can be injected into the DSA process.
}
\end{figure}
\clearpage

\begin{figure}
\vspace{-1cm}
\begin{center}
\includegraphics[scale=0.85]{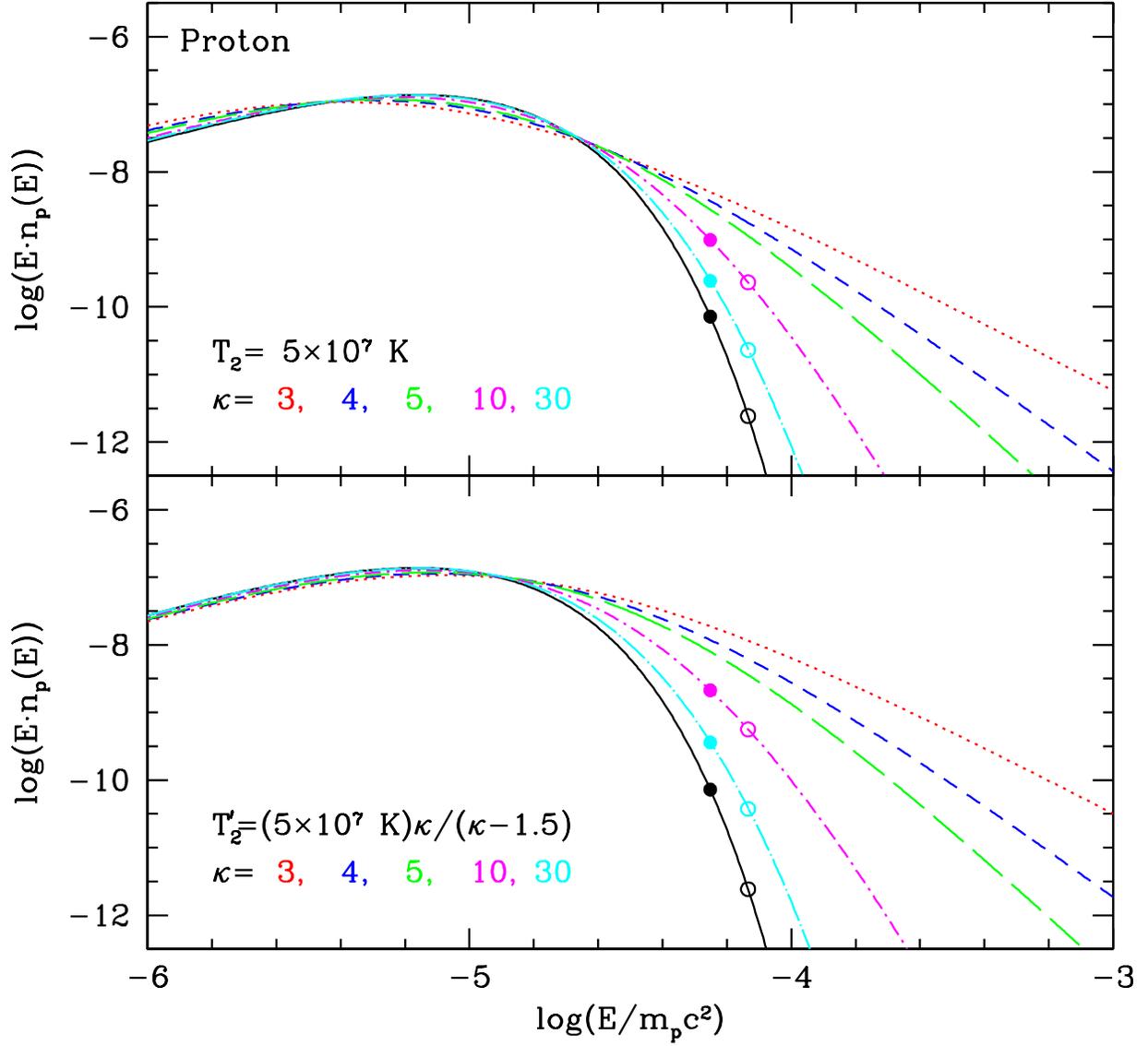}
\end{center}
\vspace{-1.0cm}
\caption{Kinetic energy spectrum for protons, $E\cdot n_p(E)$, for kinetic
temperature 
$T_2=5\times 10^7$ K (top) and $T_2^{\prime}$ given equation (\ref{Tprime})
(bottom). 
The kinetic energy, $E$, is expressed in units of $m_p^2c$
and the energy spectrum, $n_p(E)$, is plotted in units of $n_2/(m_p c^2)$.
The Maxwellian distributions are shown in (black) solid lines, while the
$\kappa$-distributions are shown in
(red) dotted, (blue) dashed, (green) long dashed, (magenta) dot-dashed, and
(cyan) dot-long dashed lines
for $\kappa=$ 3, 4, 5, 10, and 30, respectively.
Three filled and open circles indicate $3.5\ p_{\rm th,p}$ and $4\ p_{\rm
th,p}$, respectively.
}
\end{figure}
\clearpage

\begin{figure}
\vspace{-1cm}
\begin{center}
\includegraphics[scale=0.85]{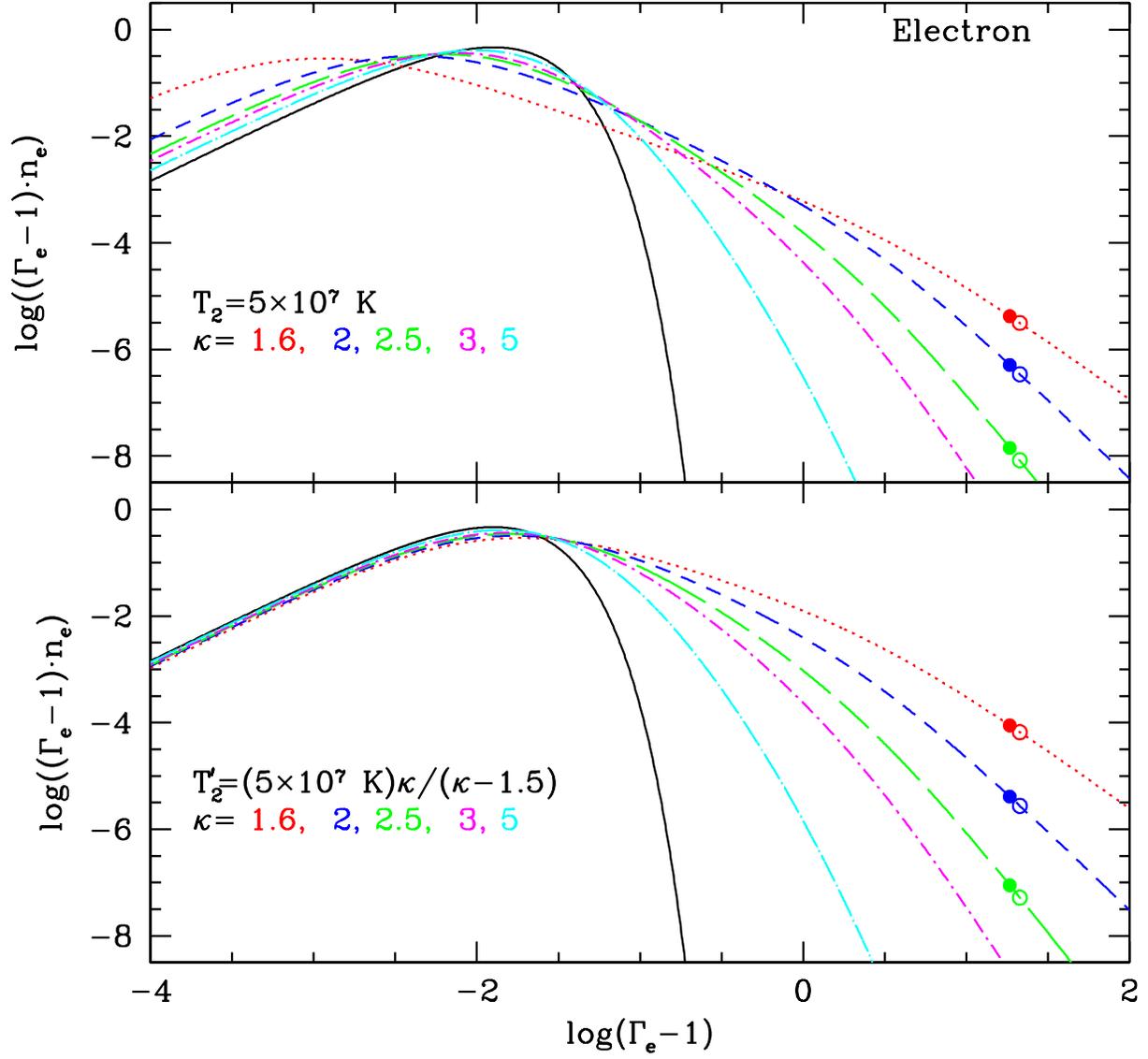}
\end{center}
\vspace{-1.0cm}
\caption{
Energy spectrum for electrons, $(\Gamma_e-1)\cdot n_e(\Gamma_e)$, for kinetic
temperature 
$T_2=5\times 10^7$ K (top) and $T_2^{\prime}$ given equation (\ref{Tprime})
(bottom),
where $\Gamma_e$ is the Lorentz factor. 
The Maxwellian distributions are shown in (black) solid lines, while the
$\kappa$-distributions are shown in
(red) dotted, (blue) dashed, (green) long dashed, (magenta) dot-dashed, and
(cyan) dot-long dashed lines
for $\kappa=$ 1.6, 2, 2.5, 3, and 5 respectively.
Three filled and open circles indicate $3.5\ p_{\rm th,p}$ and $4\ p_{\rm
th,p}$, respectively.
}
\end{figure}
\clearpage

\begin{figure}
\vspace{-1cm}
\begin{center}
\includegraphics[scale=0.85]{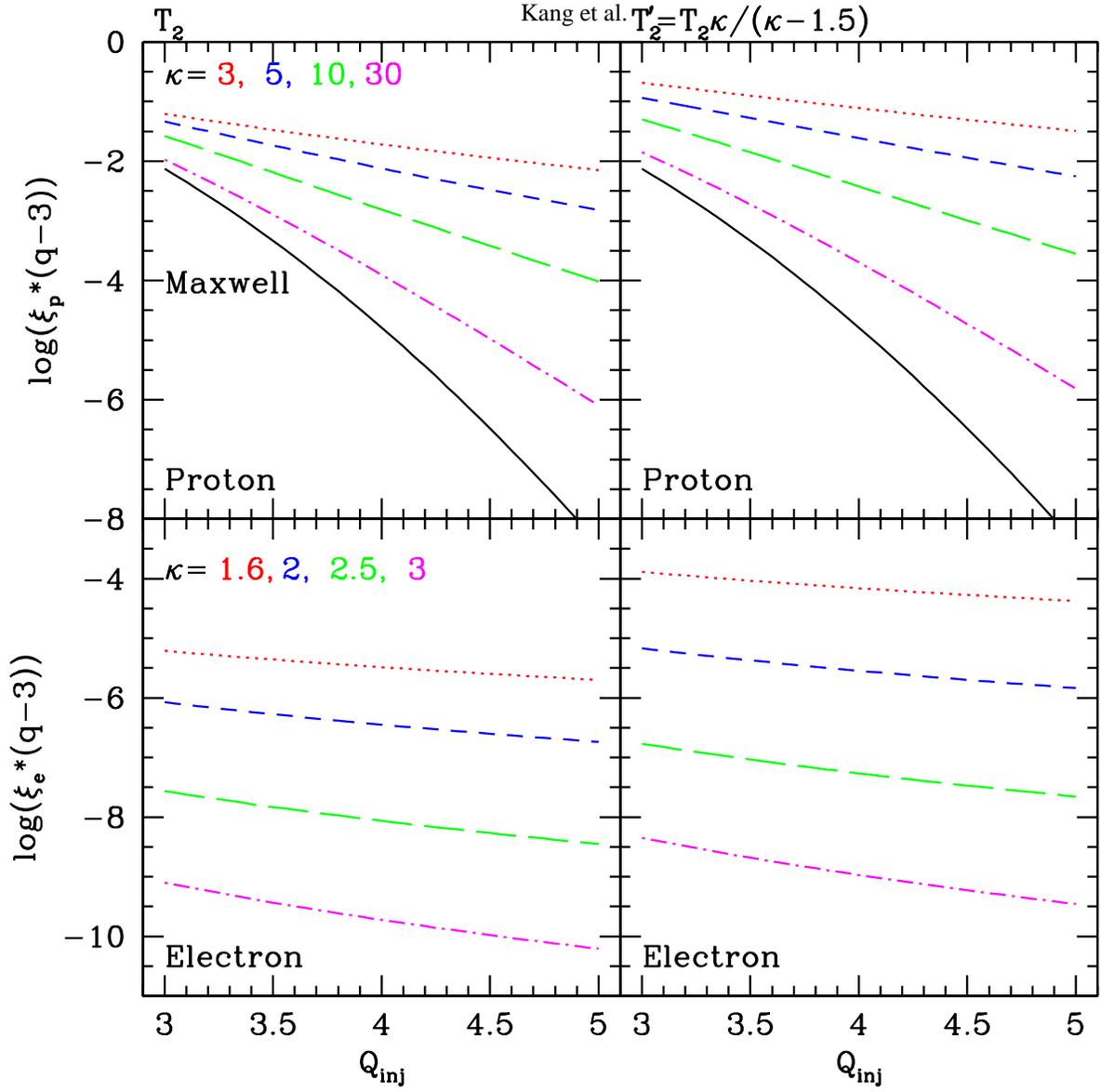}
\end{center}
\vspace{-1.0cm}
\caption{
Fraction of injected CR protons, $\xi_p\cdot(q-3)$, and electrons, $\xi_e
\cdot(q-3)$,
given in equations (\ref{xiM}) $-$ (\ref{xiK}) as a function 
$Q_{\rm inj}=p_{\rm inj}/p_{\rm th,p}$ for the
$T_2$ (left panels) and $T_2^{\prime}$ (right panels) model.
The CR spectrum is assumed to be a power-law given in equation (\ref{ftp}) for
$p\ge p_{\rm inj}$.
For protons (top panels) the Maxwellian distributions are shown in (black) solid
lines, 
while the $\kappa_p$-distributions are shown in
(red) dotted, (blue) dashed, (green) long dashed, and (magenta) dot-dashed lines
for $\kappa=$ 3, 5, 10, and 30, respectively.
For electrons (bottom panels) only the $\kappa$-distributions are shown in
(red) dotted, (blue) dashed, (green) long dashed, and (magenta) dot-dashed lines
for $\kappa=$ 1.6, 2, 2.5, and 3, respectively.
}
\end{figure}
\clearpage


\begin{thebibliography}{}

\bibitem[Amano \& Hoshino(2009)]{amano09}
Amano, T., \& Hoshino, M.
2009, \apj, 690, 244

\bibitem[Bell(1978)]{bell78}
Bell, A. R. 
1978, \mnras, 182, 147

\bibitem[Blandford \& Eichler(1987)]{blaeic87} 
Blandford, R.~D., \& Eichler, D. 1987, Phys. Rept., 154, 1

\bibitem[Brunetti \& Jones(2014)]{brunetti2014}
Brunetti, G., \& Jones, T. W.
2014, Int. J. of Modern Physics D. 23, 000

\bibitem[Caprioli \& Spitkovsky(2013)]{capri13}
Caprioli, D. \& Spitkovsky, A. 2013,
\apj, 765, L20

\bibitem[Drury(1983)]{dru83} 
Drury, L.~O'C. 
1983, Rept. Prog. Phys., 46, 973

\bibitem[Feretti et al.(2012)]{feretti12}
Feretti, L., Giovannini, G., Govoni, F., \& Murgia, M.
2012, A\&A Rev, 20, 54

\bibitem[Garat\'e \& Spitkovsky(2012)]{garat12}
Gargat\'e L. \& Spitkovsky, A. 2012, 
\apj, 744, 67

\bibitem[Giacalone(2005)]{giacalone05}
Giacalone, J.
2005, \apj, 628, L37

\bibitem[Gosling et al.(1989)]{gosling89}
Gosling, J. T., Thomsen, M. F.,\& Bame, S. J. 
1989, \jgr, 94, 10011

\bibitem[Guo \& Giacalone(2010)]{guo10}
Guo, F., \& Giacalone, J.
2010, \apj, 715, 406

\bibitem[Guo \& Giacalone(2013)]{guo13}
Guo, F., \& Giacalone, J.
2013, \apj, 773, 158

\bibitem[Hamilton \& Petrosian(1992)]{hamilton92}
Hamilton, R. J., \& Petrosian, V. 
1992, \apj, 398, 350

\bibitem[Hellberg et al.(2000)]{hellberg00}
Hellberg, M. A., Mace, R. L., Armstrong, R. J., \& Karlstad, G. 
2000, J. Plasma Phys. 64, 433

\bibitem[Hillas(2005)]{hill05}
Hillas, A. M., 2005,
Journal of Physics G, 31, R95

\bibitem[Jones \& Ellison(1991)]{jones91} 
Jones, F.C., \& Ellison, D.C. 
1991,\ssr, 58, 259

\bibitem[Kang(2011)]{kang11}
Kang, H. 2011,
J. Korean Astron. Soc., 44, 1

\bibitem[Kang(2012)]{kang12}
Kang, H. 2012,
J. Korean Astron. Soc., 45, 127

\bibitem[Kang et al.(2002)]{kjg02}
Kang, H., Jones, T. W., \& Gieseler, U.D.J. 
2002, \apj, 579, 337

\bibitem[Kang \& Ryu(2010)]{kr10}
Kang, H., \& Ryu, D. 
2010, \apj, 721, 886

\bibitem[Kang et al.(2012)]{krj12}
Kang, H., Ryu, D., \& Jones, T. W. 
2012, \apj, 756, 97

\bibitem[Leubner(2004)]{leubner04}
Leubner, M.P.
2004, Physics of Plasmas, 11, 1308

 
\bibitem[Malkov \& Drury(2001)]{maldru01}
Malkov, M. A., \& Drury, L.O'C. 2001,
Rep. Progr. Phys., 64, 429

\bibitem[Masters et al.(2013)]{masters13}
Masters, A., Stawarz, L., Fujimoto, M., et al.
2013, Nat. Phys., 9, 164 

\bibitem[Matsumoto et al.(2013)]{matsumoto13}
Matsumoto, Y., Amano, T., \& Hoshino, M.
2013, \prl, 111, 215003

\bibitem[Mendoza \& Bautista(2014)]{mendoza14}
Mendoza, C. \& Bautista, M. A.
2014, \apj, 785, 91

\bibitem[Neergaard-Parker \& Zank(2012)]{parker12}
Neergaard-Parker, L., \& Zank, G.P.
2012, \apj, 757, 97

\bibitem[Neergaard-Parker et al.(2014)]{parker14}
Neergaard-Parker, L., Zank, G.P., \& Hu, Q.
2014, \apj, 782, 52

\bibitem[Nicholls et al.(2012)]{nicholls12}
Nicholls, D.C., Dopita, M.A., \& Sutherland, R. S. 
2012, \apj, 752, 148

\bibitem[Oka et al.(2006)]{oka06}
Oka, M., Terasawa, T., Seki, Y., et al. 
2006, \grl, 33, L24104


\bibitem[Petrosian(2012)]{petrosian12}
Petrosian, V. 
2012, \ssr, 173, 535

\bibitem[Petrosian \& East(2008)]{petrosian08}
Petrosian, V., \& East, W. E. 
2008, \apj, 682, 175

\bibitem[Petrosian \& Lui(2004)]{petrosian04}
Petrosian, V., \& Lui, S.
2004, \apj, 610, 550

\bibitem[Pierrard \& Lazar(2010)]{pierrard10}
Pierrard, V., \& Lazar, M. 2010, Sol. Phys., 265, 153


\bibitem[Pinzke et al.(2013)]{pop13}
Pinzke, A., Oh, S. P., \& Pfrommer, C. 
2013 \mnras, 435, 1061 

\bibitem[Pryadko \& Petrosian(1997)]{pryadko97}
Pryadko, J. M., \& Petrosian, V. 
1997, \apj, 482, 774

\bibitem[Quest(1988)]{quest88}
Quest, K.B.
1988, \jgr, 93, 9649

\bibitem[Reynolds et al.(2012)]{reynolds12}
Reynolds, S. P., Gaensler, B. M., \& Bocchino, F. 
2012, \ssr, 166, 231

\bibitem[Riquelme \& Spitkovsky(2011)]{riqu11}
Riquelme, M. A., \& Spitkovsky, A. 
2011, \apj, 733, 63

\bibitem[Ryu \etal (2003)]{ryuetal03}
Ryu, D., Kang, H., Hallman, E., \& Jones, T. W. 2003,
\apj, 593, 599

\bibitem[Schure et al.(2012)]{schure12}
Schure, K. M., Bell, A. R, Drury, L. O'C., \&. Bykov, A. M.
2012, \ssr, 173, 491

\bibitem[Shimada et al.(1999)]{shimada99}
Shimada, N., Terasawa, T., Hoshino, M., et al.
1999, Ap\&SS, 264, 481

\bibitem[Simnett et al.(2005)]{simnett05}
Simnett, G.M., Sakai, J.-I., \& Forsyth, R.J.
2005, \aap, 440, 759

\bibitem[Vasyliunas(1968)]{vasyliunas68}
Vasyliunas, V. M.
1968, \jgr, 73, 2839

\bibitem[van Weeren et al.(2010)]{wrbh10}
van Weeren, R., R\"ottgering, H. J. A., Br\"uggen, M., \& Hoeft, M.
2010, Science, 330, 347

\bibitem[Wilson et al.(2009)]{wilson09} 
Wilson, L. B., III, Cattell, C. A., Kellogg, P. J. et al.
2009, \jgr, 114, A10106

\bibitem[Wilson et al.(2012)]{wilson12} 
Wilson, L. B., III, Koval, A., Szabo, A. et al.
2012, \grl, 39, L08109

\bibitem[Zank et al.(2006)]{zank06}
Zank, G. P., Li, G., Florinski, V., et al. 
2006, \jgr, 111, A06108

\bibitem[Zank et al.(2007)]{zank07} 
Zank, G.P, Li, G., \& Verkhoglyadova, O. 
2007, \ssr, 130, 255 

\end{thebibliography}
\end{document}